\documentclass[10pt, conference, letterpaper]{IEEEtran}

\usepackage{graphicx}
\graphicspath{{../pdf/}{../jpeg/}{../PNG/}}
\usepackage{amsmath}
\DeclareGraphicsExtensions{.pdf,.jpeg,.png}

\usepackage{mathtools}
\usepackage{amsmath,amssymb}
\usepackage{amsfonts}
\usepackage{braket}
\usepackage{bold-extra}

\usepackage[tableposition=top]{caption} 
\usepackage{lipsum}
\usepackage{cite}                                                          
\usepackage{stackengine}
\usepackage[font=footnotesize]{caption}

\usepackage{eqparbox}
\usepackage{subcaption}
\usepackage{url}
\usepackage[tableposition=top]{caption} 
\usepackage{lipsum}

\usepackage{array}
\usepackage{mdwmath}
\usepackage{mdwtab}

\usepackage{bibentry}  
\usepackage{xcolor,soul,framed} 
\usepackage{multicol}

\usepackage{epsfig} 
\usepackage{mathptmx} 
\usepackage{times} 
\usepackage{tikz}

\usetikzlibrary{backgrounds,fit,decorations.pathreplacing,calc}

\usepackage{qcircuit}

\title{\LARGE \bf
An Improved Implementation Approach for Quantum Phase Estimation on Quantum Computers
}

\author{Hamed Mohammadbagherpoor$^{1}$, Young-Hyun Oh$^{2*}$, Patrick Dreher$^{3}$, Anand Singh$^{2}$, Xianqing Yu$^{2}$, Andy J. Rindos$^{2}$\\  
$^{1,3}$Department of ECE $\&$ CSC,
\\
North Carolina State University, Raleigh, NC 27606, USA\\
$^{2}$Cloud and Cognitive Software, IBM, Durham, NC 27703, USA 
\\
Email: ohy@us.ibm.com$^{*}$
}

\usepackage{lastpage}
\begin{document}

\maketitle

\begin{abstract}
Quantum phase estimation (QPE) is one of the core algorithms for quantum computing. It has been extensively studied and applied in a variety of quantum applications such as the Shor's factoring algorithm, quantum sampling algorithms and the calculation of the eigenvalues of unitary matrices. The QPE algorithm has been combined with Kitaev's algorithm and the inverse quantum Fourier transform (IQFT) which are utilized as a fundamental component of such quantum algorithms. In this paper, we explore the computational challenges of implementing QPE algorithms on noisy intermediate-scale quantum (NISQ) machines using the IBM Q Experience (e.g., the IBMQX4, 5-qubit quantum computing hardware platform).  Our experimental results indicate that the accuracy of finding the phase using these QPE algorithms is severely constrained by the NISQ computer's physical characteristics such as coherence time and error rates. To mitigate these physical limitations, we propose implementing a modified solution by reducing the number of controlled rotation gates and phase shift operations, thereby increasing the accuracy of the finding phase in near-term quantum computers. 
\end{abstract}

\section{INTRODUCTION}

The first generation of noisy intermediate-scale quantum (NISQ)~\cite{c23} computers now provides a framework for re-formulating algorithms originally optimized for digital computers into a form suitable for the new quantum computing hardware platforms. These re-formulations hold the promise of potentially being able to solve particular problems exponentially faster than classical computers and also to explore regions that are inaccessible using even the most powerful digital high performance computers.  

A key difference between an algorithm that is formulated for a digital computer versus a quantum computer is that digital computations are modeled on a Load-Run-Read cycle while quantum computers operate on a Prepare-Evolve-Measure cycle. The information flow for digital algorithms assumes that input data in digital bit format is inserted into the system, the program runs and then the output of the program is read. However, in quantum computers the qubit states are prepared as the input, manipulation of the input states is done using the operators and then the results are measured~\cite{c21}. As part of the design for information flow is a quantum computer developer will also incorporate the quantum mechanical properties of both of superposition and entanglement of the qubits~\cite{c12,c24} in order to strive for a quantum advantage over their digital counterparts.  

Quantum phase estimation (QPE) is the critical building block for various quantum algorithms. In QPE the main objective of quantum phase estimation is to determine the eigenvalues of an unitary matrix with an unchanged eigenvector. This technique was described by Kitaev~\cite{c26}. This procedure is a critical component in QC algorithm development for quantitative finance as well as mathematics such as Shor's algorithm for factoring the prime numbers, Grover's algorithm to search~\cite{c1,c2,c3,c4,c5,c6,c11}, cryptography, physics and quantum chemistry. Today there is an active research program to approximate, parallelize, and decompose quantum phase estimation related to algorithms~\cite{c27, c28, c29}.

However, implementing quantum algorithms on near-term quantum computers are severely constrained by low number of qubits, low reliability and high variability of quantum computers' physical characteristics. For example, the largest number factored by actual quantum computer is the number $143$ which was implemented on a dipolar coupling NMR System by applying adiabatic quantum computation~\cite{c20}. In addition, Shor's algorithm for factoring 15 (i.e., 3*5) on a nuclear magnetic resonance (NMR) computer is presented in~\cite{c18} and the number 21 is factored by implementing qubit recycling in a photonic circuit~\cite{c19}. Although the number of qubits is small, these experimental approaches will be considerably valuable when we can take full advantage of quantum supremacy in near future. 

There are two main approaches that are used to implement quantum phase estimation. The first approach is to extract the phase information by applying the classical post processing computation after utilizing quantum gate operations as known as Kitaev's algorithm~\cite{c7,c8}. Because Kitaev's algorithm requires some classical post processing after performing Hadamard operations, it is necessary to run a minimal number of trials to obtain the phase $k^{th}$-bit position with constant success probability. The second approach is to find the phase information in which the phase is estimated by applying inverse quantum Fourier transform (IQFT)~\cite{c9,c10,c13}. However, IQFT approach requires a large number of rotation gates for precision digits to obtain more accurate phase information. Without loss of generality, more rotation gates can cause more readout errors from implementation of IQFT algorithms on near-term quantum computers. Thus, it is critical to minimize depth and controlled-rotation gates to increase the accuracy of finding the phase information. 

There have been several experimental hardware platforms constructed to test some of these QPE implementations. An experimental phase estimation based on quantum Fourier transform was implemented on a three-bit nuclear magnetic resonance (NMR) processor~\cite{c16} but it only used to estimate the eigenvalues of one-bit Grover operators. An implementation of phase estimation algorithm on an ion-trapped quantum computer was proposed to find the eigenstates of the system~\cite{c17}. Lloyd et al. have shown that quantum computers can speed up of some linear algebraic based machine learning algorithms by applying quantum phase estimation technique such as principle component analysis (PCA), support vector machine (SVM), and K-means algorithms~\cite{c14,c15}. 

In this paper, we have implemented various quantum phase estimation algorithms using both the Qiskit Aer quantum simulator~\cite{c30} for the theoretical results and the IBM Q Experience~\cite{c22} for experimental results from the perspective of NISQ physical limitations. The experimental results show that the accuracy of finding the correct phase decreases as the number of qubits and quantum operations increase. To mitigate the problem, we propose modified solutions of these QPE algorithms by minimizing the number of control gates and phase shift operators.

This paper is categorized as follows. Section \ref{sect-2} describes the basic quantum operations and various phase estimation algorithms such as Kitaev's algorithm, the iterative algorithm to estimate the phase, Lloyd algorithm for phase estimation based on inverse quantum Fourier transform (IQFT), and the constant precision algorithm. In section \ref{sect-3}, the simulation and experimental results for each method are provided and compared. Finally, Section \ref{sect-4} summarizes the results and conclusions from this work.

\begin{figure}[t]
 \begin{center}
 \Large 
\[ \Qcircuit @C=1cm @R=1cm {
         \lstick{\ket{0}}    & \gate{H} &  \ctrl{1} &\rstick{\frac{1}{\sqrt{2}}(\ket{0}+e^{2\pi i \varphi}\ket{1})}\qw\\
         \lstick{\ket{\psi}} & \qw     &  \gate{U} & \rstick{\ket{\psi}}\qw & \push{\rule{3.5em}{0em}}
     } \]
\caption{Quantum circuit for transforming the states}
  \label{Ucontrolled}
   \end{center}
\vspace{-1.5em}
\end{figure}
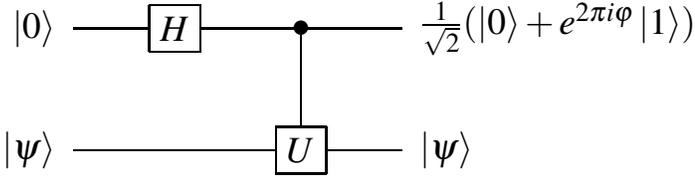


\section{Phase estimation}
\label{sect-2}

Phase estimation is a technique that is used to estimate the eigenvalues $\ket{\lambda}$ of a unitary matrix $U$ with its known eigenvector $\ket{\psi_\lambda}$~\cite{c12},

\begin{equation}
   U \ket{\psi_\lambda} = \lambda \ket{\psi_\lambda},
   \\
\end{equation}

\noindent where the eigenvalues of the unitary matrix are $\lambda =e^{2\pi i \varphi_n}$. The phase of the unitary matrix can be written as $\varphi_n = 0.x_1x_2x_3...x_n$ where $n$ is the number of qubits used for phase estimation. The estimated variable $(\hat{\varphi})$ can be expressed as a binary representation,

\begin{equation}
   \hat{\varphi} = \frac{x_1}{2^1}+\frac{x_2}{2^2}+\frac{x_3}{2^3}+\cdot \cdot\cdot+\frac{x_n}{2^n} 
   \\
\end{equation}

Fig.\ref{Ucontrolled} illustrates a quantum computing circuit that incorporates this problem of phase determination. The circuit consists of one qubit and an eigenstate, a Hadamard gate (H) and a rotation gate (U). The output of the circuit contains the phase $\frac{1}{\sqrt{2}}\big(\ket{0}+e^{i2\pi\varphi}\ket{1}\big)$ described in the top of the Fig.\ref{Ucontrolled}.  

\begin{figure}[t]
 \begin{center}
 \Large{} 
 \[\Qcircuit @C=.5cm @R=.8cm {
         \push{\rule{.5em}{0em}} & \lstick{\ket{0}}    & \gate{H} & \gate{K} &  \ctrl{1} & \gate{H} & \meter  \\
         \push{\rule{.5em}{0em}} & \lstick{\ket{\psi}} & \qw      & \qw      &  \gate{U^{2^{k-1}}} &  \qw & \rstick{\ket{\psi}}\qw  
     }\] 
\caption{Controlled U circuit}
  \label{Kitaev}
   \end{center}
\vspace{-1.5em}
\end{figure}
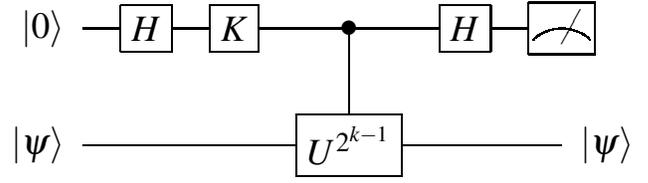


The goal of phase estimation is to find the eigenvalues of an unitary matrix and then apply these eigenvalues to estimate the unknown phase of the unitary operator. However, with only this information available from the circuit output it is impossible to find the correct phase due to the superposition state on the value.  

There are different methods used to calculate the phase for an unitary matrix expressed in this circuit construction. Section \ref{sect-kitaev} describes sequential post processing techniques to calculate the unknown phase. Section \ref{I-QPE} introduces an iterative technique to implement the Kiteav's algorithm with higher accuracy. Section \ref{inv-QFT} describes applying the inverse quantum Fourier transform (IQFT) to derive the unknown phase information and Section \ref{constant-P} discusses the arbitrary precision QPE that reduces the number of shift operators for phase estimation thereby decreasing the depth of the quantum circuits. These techniques along with their simulated and experimentally implemented results will be discussed in detail below.

\subsection{Kitaev's algorithm }
\label{sect-kitaev}

Kitaev's algorithm is the first algorithm that was introduced to estimate the phase of an unitary matrix. In this technique a set of Hadamard gates are applied to the input qubits. The outputs of the Hadamard gates connected with the controlled-$U^{2^{k-1}}$ result in an output represented by a phase shift operator. Applying a controlled-$U$ operator $k$ times transforms the control qubit to $\frac{1}{\sqrt{2}}\big(\ket{0}+e^{-i2\pi\varphi_k 2^{k-1}}\ket{1}\big)$. At each test phase $\varphi_k = 2^{k-1}\varphi$ can be calculated. By doing the test $k$ times and measuring the output of each test the set of values ${\varphi, 2\varphi, \cdot\cdot\cdot, 2^{k-1}\varphi}$ can be achieved. These measurements are used to estimate the phase of the unitary matrix.

Fig.\ref{Kitaev} shows the circuit that performs the phase estimation. The operation $K$ can be used to manipulate the qubit phase and provides more information about the phase of the system. Considering a $2x2$ identity matrix $I_2$ and setting, 
\begin{equation}
K = I_2 =
\begin{pmatrix}
    1       & 0 \\
    0       & 1 \\
\end{pmatrix}
\end{equation}

\noindent the mathematical manipulations of the qubits and the introduction of the phases can be seen in Eq.~\ref{Kitaev_eq}.

\begin{equation}
\label{Kitaev_eq}
\begin{array}{l}

    \ket{0}\ket{\psi_\lambda} \enspace\>  \xrightarrow{H\bigotimes I} \qquad \frac{1}{\sqrt{2}}(\ket{0}+\ket{1})\ket{\psi_\lambda}
    \\
    \\
    \>\qquad\qquad \xrightarrow{C-U_k} \qquad \frac{1}{\sqrt{2}}(\ket{0}+\ket{1})\bigotimes U_k \ket{\psi_\lambda}
    \\
    \\
    \qquad\qquad    = \frac{1}{\sqrt{2}}(\ket{0}\ket{\psi_\lambda}+e^{2\pi i\varphi_k}\ket{1}\ket{\psi_\lambda}) 
    \\
    \\
    \xrightarrow{H\bigotimes I}  
    \quad \frac{1}{\sqrt{2}}\frac{\big(\ket{0}+
    \ket{1}\big)}{\sqrt{2}}\ket{\psi_\lambda} +\frac{e^{2\pi i\varphi_k}}{\sqrt{2}}\frac{\big(\ket{0}-\ket{1}\big)}{\sqrt{2}}\ket{\psi_\lambda} 
    \\
    \\ 
    \qquad\quad = \frac{1}{2}\bigg(\big(1+e^{2\pi i\varphi_k}\big)\ket{0} + \big(1-e^{2\pi i\varphi_k}\big)\ket{1}\bigg)\ket{\psi_\lambda} 

\end{array}
\end{equation}

Based on the calculations from Eq.~\ref{Kitaev_eq}, the probability of measuring $\ket{0}$ and $\ket{1}$ will be,

\begin{equation}
\label{prob_cosine}
   P(0|k) = \frac{1+cos(2\pi\varphi_k)}{2}, \qquad P(1|k) = \frac{1-cos(2\pi\varphi_k)}{2}
\end{equation}
The quantity $\varphi_k$ can be obtained more precisely by applying more trials. However, based on the data from Eq.~\ref{Kitaev_eq} we cannot distinguish between $\varphi_k$ and $-\varphi_k$. Another circuit is required to provide more information about the phase of the unitary matrix to distinguish between $\varphi_k$ and $-\varphi_k$. 

By considering the combination of the results from $K=I_2$ and $K=S$ the actual value of the phase can be determined. 
\begin{equation}
K = S =
\begin{pmatrix}
    1       & 0 \\
    0       & i
\end{pmatrix}
\end{equation}
Eq. \ref{Kitaev_S_part1} illustrates that quantum circuit provides the following  transformation if the $K=S$ gate is applied to the circuit.  

\begin{equation}
\label{Kitaev_S_part1}
\begin{array}{l}

   \ket{0}\ket{\psi_\lambda} \enspace\> \xrightarrow{H\bigotimes I} \qquad \frac{1}{\sqrt{2}}(\ket{0}+\ket{1})\ket{\psi_\lambda}\\
   \\
    \>\qquad\qquad \xrightarrow{S} \qquad \enspace\enspace\>\frac{1}{\sqrt{2}}(\ket{0}+i\ket{1}) \ket{\psi_\lambda}
    \\
    \\
   \qquad\qquad \xrightarrow{C-U_k} \qquad \frac{1}{\sqrt{2}}(\ket{0}+i\ket{1})\bigotimes U_k \ket{\psi_\lambda}\\
   \\
   \qquad\qquad= \frac{1}{\sqrt{2}}(\ket{0}\ket{\psi_\lambda}+i e^{2\pi i\varphi_k}\ket{1}\ket{\psi_\lambda})\\ 
   \\ 
   \xrightarrow{H\bigotimes I} 
   \frac{1}{\sqrt{2}}\frac{\big(\ket{0}+i \ket{1}\big)}{\sqrt{2}}\ket{\psi_\lambda} +i \frac{e^{2\pi i\varphi_k}}{\sqrt{2}}\frac{\big(\ket{0}-i\ket{1}\big)}{\sqrt{2}}\ket{\psi_\lambda}\\
   \\
   = \frac{1}{2}\bigg(\big(1+ie^{2\pi i\varphi_k}\big)\ket{0} + \big(1-ie^{2\pi i\varphi_k}\big)\ket{1}\bigg)\ket{\psi_\lambda} \\
    \\
   = \frac{1}{2}\bigg(\big(1+e^{2\pi i\varphi_k+\frac{\pi}{2}}\big)\ket{0} + \big(1-ie^{2\pi      i\varphi_k+\frac{\pi}{2}}\big)\ket{1}\bigg)\ket{\psi_\lambda}
   
\end{array}
\end{equation} 
    


Based on the calculations from Eq.~\ref{Kitaev_S_part1}, the probability of measuring $\ket{0}$ and $\ket{1}$ will be,

\begin{equation}
\label{prob_sine}
   P(0|k) = \frac{1-sin(2\pi\varphi_k)}{2}, \qquad P(1|k) = \frac{1+sin(2\pi\varphi_k)}{2}
\end{equation}


\noindent Eq.\ref{prob_sine} provides the additional information needed to determine the correct phase of the unitary matrix.  In each test the probabilities of being zero or one in $t$ trials are measured. By using the results from Eq.\ref{prob_cosine} and Eq.\ref{prob_sine} the estimation of $cos(2\pi\varphi_k)$, $sin(2\pi\varphi_k)$, and the phase $(\hat{\varphi})$ can be calculated by  
\begin{equation}
\label{tangent}
   \hat{\varphi_k} = \frac{1}{2\pi}\;tan^{-1} \bigg(\frac{C_k}{S_k}\bigg)
\end{equation}
where $C_k$ and $S_k$ are the estimation of $cos(2\pi\varphi_k)$ and $sin(2\pi\varphi_k)$ respectively.

In Kitaev's algorithm post processing calculation is required to estimate the value of the phase. Estimating of the phase within $m$ bits of accuracy requires to increase the number of trials. $O\big(\frac{log(1-\delta)}{\epsilon}\big)$ samples are required to estimate within $\epsilon$ with probability of $1-\delta$.

\begin{table}[b]
 \begin{center}
  \includegraphics[width=3.5in]{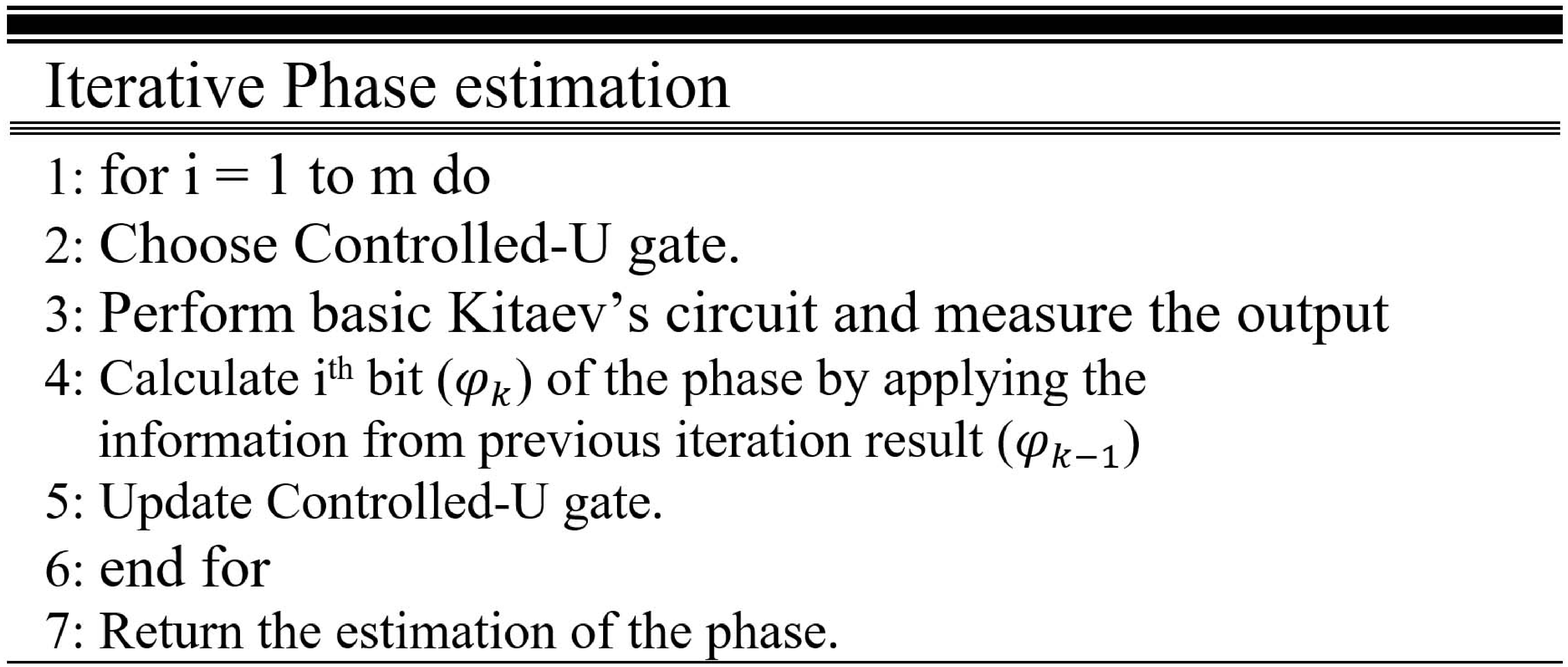}
  \caption{Iterative quantum phase estimation}
  \label{Iterative_PE}
  \end{center}
\vspace{-1.5em}
\end{table}

\subsection{Iterative quantum phase estimation}
\label{I-QPE}

In our experimental implementation, increasing the number of gates to estimate the phase with higher accuracy increases the convergence error so that the ability to approach the correct answer degrades. This section describes the iterative technique composed of the Kitaev's algorithm as a main component that estimates the phase with high accuracy in finite set of iterations. In order to estimate the phase, however, the iterative Kitaev's algorithm requires not only to run with sufficient number of shots and measurements but also to conduct post-processing calculation to determine the phase. Svore et al. introduces a fast phase estimation algorithm which considers interference across multiple qubits and asymptotically improve in runtime with less number of measurements and lower circuit width and depth~\cite{c31}. Another approach has been discussed in~\cite{c32} in which an adaptive algorithm based on Bayes' rule is provided to estimate the uncertainty of the phase using the experimental data. The probability distribution is updated by Bayes' rule by analyzing the previous set of experiment.  

Table \ref{Iterative_PE} shows the general iterative Kitaev's algorithm that helps to find the unknown phase of the system with $m$ bits of accuracy. One Hadamard gate is used to perform the superposition. A controlled-U gate is then applied to the output of the Hadamard. The next step applies another Hadamard gate and then performs the final measurement for that iteration. 
In the next iteration the order of controlled-U gates is updated and the result from the previous measurement is applied to the circuit to estimate the new bit. This technique is repeated $m$ times to estimate the phase with \textit{m} bits of accuracy. In this method, each iteration of information from the previous iterations is used to estimate the next bit of the phase.

\begin{figure}[t]
 \begin{center}
 \large{} 
 \[\Qcircuit @C=.2cm @R=.2cm @!R {
         \lstick{\ket{0}}       & \gate{H} & \qw & \qw & \qw &\qw & & & \ctrl{3} &\qw &  \multigate{2}{QFT^\dag} & \rstick{\ket{x_n}}\qw\\
         \lstick{\ket{0}}       & \gate{H} & \qw & \qw & \ctrl{2} &\qw & &  &\qw &\qw & \ghost{QFT^\dag} &\rstick{\ket{x_{n-1}}}\qw \\
         \lstick{\ket{0}}       & \gate{H} & \ctrl{1} & \qw & \qw & \qw & & &\qw &\qw& \ghost{QFT^\dag} & \rstick{\ket{x_1}}\qw \\
         \lstick{\ket{\psi}}  & \qw  & \gate{U^{2^{0}}}  & \qw  & \gate{U^{2^{1}}} & \qw  & & & \gate{U^{2^{k-1}}} & \qw& \qw& \qw& \rstick{\ket{\psi}}\qw\\ 
     }\]
\caption{Quantum phase estimation based on inverse quantum Fourier transform}
  \label{QFT}
   \end{center}
\vspace{-1.5em}
\end{figure}
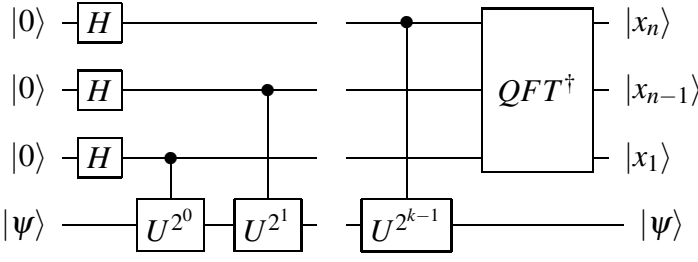

\subsection{Phase estimation based on inverse QFT}
\label{inv-QFT}

One of the common methods used to implement the QPE algorithm is based on inverse QFT. The general view of this method has been shown in Fig.\ref{QFT}. In this method two stages are required for phase estimation. The first stage starts with n-qubits initialized at $\ket{0}$ and prepares the state $\ket{\psi}$. The second stage uses inverse quantum Fourier transform operation to estimate the binary digits of the phase. 

The mathematical equations of the first stage are given by Eq. \ref{Kitaev_eq_sine}.

\begin{equation}
\label{Kitaev_eq_sine}
\begin{array}{l}
   \frac{1}{\sqrt{2}}\big(\ket{0}+e^{2\pi i 2^{n-1}\varphi}\ket{1}\big) 
   \frac{1}{\sqrt{2}}\big(\ket{0}+e^{2\pi i 2^{n-2}\varphi}\ket{1}\big)\cdot\cdot\cdot \\ 
   \\ 
   \frac{1}{\sqrt{2}}\big(\ket{0}+e^{2\pi i \varphi}\ket{1}\big) = \frac{1}{2^{n/2}} \sum_{k=0}^{2^{n-1}}{e^{2\pi i \frac{\varphi k}{2^n}} \ket{k}}
   \\ 
\end{array}
\end{equation}

Considering $\varphi = x/2^n$ where $x = \sum_{i=0}^{n-1}{2^i x_i}$ produces the Eq. \ref{QFT_2}

\begin{equation}
\label{QFT_2}
\begin{array}{l}
   \frac{1}{\sqrt{2}}\big(\ket{0}+e^{2\pi i 0.x_n}\ket{1}\big) 
   \frac{1}{\sqrt{2}}\big(\ket{0}+e^{2\pi i 0.x_{n-1}x_n\varphi}\ket{1}\big)\cdot\cdot\cdot \\
   \\ 
   \frac{1}{\sqrt{2}}\big(\ket{0}+e^{2\pi i 0.x_1x_2...x_n}\ket{1}\big) = \frac{1}{2^{n/2}} \sum_{k=0}^{2^{n-1}}{e^{2\pi i \frac{\varphi k}{2^n}} \ket{k}}
   \\
\end{array}   
\end{equation}

\begin{figure}[b]
 \begin{center}
 \Large{} 
 \[\Qcircuit @C=1cm @R=1cm @!R {
         \push{\rule{3em}{0em}} & \lstick{\frac{1}{\sqrt{2}}(\ket{0}+e^{2\pi i 0.x_1}\ket{1})}    & \gate{H} & \rstick{\ket{x_1}}\qw 
     }\]
\caption{One bit phase estimation quantum circuit}
  \label{One_digit}
   \end{center}
\end{figure}


\begin{figure}[t]
 \begin{center}
 \Large{} 
 \[\Qcircuit @C=.3cm @R=.5cm  {
         \push{\rule{4em}{0em}} & \lstick{\frac{1}{\sqrt{2}}(\ket{0}+e^{2\pi i 0.x_2}\ket{1})}    & \gate{H} &  \ctrl{1} & \qw & \rstick{\ket{x_2}}\qw \\
         \push{\rule{7em}{0em}} & \lstick{\frac{1}{\sqrt{2}}(\ket{0}+e^{2\pi i 0.x_1x_2}\ket{1})} &  \qw &  \gate{R_2 ^\dag}  & \gate{H} & \rstick{\ket{x_1}}\qw 
     }\]
\caption{3-Qubits inverse quantum Fourier Transform (IQFT)}
  \label{3_QbitQFT}
   \end{center}
\vspace{-1.5em}
\end{figure}

As can be seen from Fig.\ref{QFT} and Eq. \ref{Kitaev_eq_sine} the outputs from the first stage (phase kick-back) are the input of inverse QFT. By applying controlled-$U^{2^{n-1}}$ there will phase kick back to prepare the states. Also, the output of the first stage is exactly the quantum Fourier transform of $\varphi$. By applying the inverse QFT we can recover the unknown phase. In order to analyze this method two different phase estimation circuits with different accuracy have been considered.


\textbf{Case1:}
Starting with $\varphi$ = $0.x_1$, as shown in the circuit in Fig. \ref{One_digit} and applying Hadamard gate to the initial state $\ket{0}$ produces the Eq. \ref{Cas1_onedigit}

\begin{equation}
\label{Cas1_onedigit}
\begin{array}{l}
   \ket{0}\enspace\>\xrightarrow{H}  \qquad \frac{1}{\sqrt{2}}\big(\ket{0}+ \ket{1}\big)\\
\\ 
\qquad \xrightarrow{U}  \qquad \frac{1}{\sqrt{2}}\big(\ket{0}+e^{2\pi i\varphi}\ket{1}\big)
    \\
    \\ 
\qquad \xrightarrow{H}     \qquad \frac{1}{2}\big(1+e^{2\pi i\varphi}\big)\ket{0} + \big(1-e^{2\pi i\varphi}\big)\ket{1} 
    \\
    \\ 
\qquad\qquad = \frac{1}{2}\big(1+e^{2\pi i 0.x_1}\big)\ket{0} + \big(1-e^{2\pi i 0.x_1}\big)\ket{1} \\

\end{array}
\end{equation} 

\noindent Calculating the probability from Eq. \ref{Cas1_onedigit} produces Eq. \ref{Probab_case1}.

\begin{equation}\label{Probab_case1}
\begin{array}{l}
  P(\ket{0}) = \frac{1+cos(2\pi0.x_1)}{2}, \qquad
  P(\ket{1}) = \frac{1-cos(2\pi0.x_1)}{2}    
 \\   
\end{array}
\end{equation} 

\noindent Based on the result from Eq. \ref{Probab_case1}, if $x_1=0$, then the probability of $\ket{0}$ is $1$ [i.e.\ $P(\ket{0})=1$] and if $x_1=1$, then the probability of $\ket{1}$ is $1$ [i.e.\ $P(\ket{1})=1$]. The conclusion that is inferred from this case is that the phase is considered as one bit and only one Hadamard gate is required to extract $x_1$.

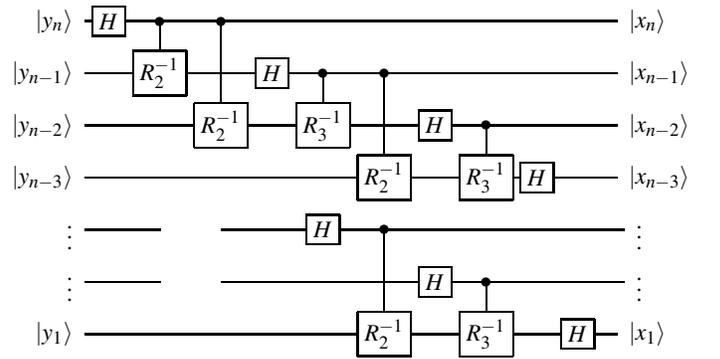
\begin{figure}[b]
 \begin{center}
 \small{} 
 \[\Qcircuit @C=.1cm @R=.1cm @!R {
         \push{\rule{2em}{0em}}&\lstick{\ket{y_n}}       & \gate{H}  & \ctrl{1} & \ctrl{2} &\qw  &\qw  &\qw &\qw &\qw &\qw&\qw&\qw&\qw  &\rstick{\ket{x_n}}\qw\\
         \push{\rule{2em}{0em}}&\lstick{\ket{y_{n-1}}}   & \qw       & \gate{R_2^{-1}} &\qw & \gate{H} & \ctrl{1} & \ctrl{2} &\qw  &\qw&\qw&\qw &\qw&\qw &\rstick{\ket{x_{n-1}}}\qw \\
         \push{\rule{2em}{0em}}&\lstick{\ket{y_{n-2}}}   & \qw       & \qw  & \gate{R_2^{-1}} & \qw & \gate{R_3^{-1}} &\qw & \gate{H} & \ctrl{1} &\qw&\qw&\qw &\qw& \rstick{\ket{x_{n-2}}}\qw \\
         \push{\rule{2em}{0em}}&\lstick{\ket{y_{n-3}}}  &\qw  & \qw & \qw & \qw  & \qw & \gate{R_2^{-1}} & \qw & \gate{R_3^{-1}} & \gate{H} &\qw&\qw &\qw& \rstick{\ket{x_{n-3}}}\qw \\
         \push{\rule{2em}{0em}}&\lstick{\vdots}& \qw & \qw && \qw  & \gate{H}  & \ctrl{2} & \qw & \qw  & \qw & \qw & \qw &\qw &\qw& \rstick{\vdots}\qw\\
         \push{\rule{2em}{0em}} &\lstick{\vdots}& \qw & \qw && \qw  & \qw & \qw & \gate{H} &  \ctrl{1}  & \qw & \qw&\qw &\qw&\qw &\rstick{\vdots}\qw\\
         \push{\rule{2em}{0em}}&\lstick{\ket{y_{1}}}   & \qw & \qw & \qw& \qw&\qw & \gate{R_2^{-1}} &  \qw & \gate{R_3^{-1}} &\qw & \gate{H}  &\qw &\qw& \rstick{\ket{x_1}}\qw \\
     }\]
\caption{QPE with arbitrary constant precision phase shift operators}
  \label{ACQPE}
   \end{center}
\end{figure}


\textbf{Case2:}
starting with $\varphi$ = $0.x_1x_2$, as shown in the circuit in Fig.\ref{3_QbitQFT} and applying inverse QFT, the unknown phase can be derived. The second digit ($x_2$) can be extracted by applying one Hadarmard gate, the same as the Case 1 described above. In order to extract the first digit ($x_1$), a controlled-rotation gate $R_2$ is required to remove the impact of the $x_2$.  This operation converts the result to case 1 and with the insertion of one Hadamard gate to estimate $x_1$, as Eq.\ref{Cas2_twodigit}

\begin{equation}\label{Cas2_twodigit}
\begin{array}{l}

\frac{1}{\sqrt{2}}\big(\ket{0}+e^{2\pi i 0.x_1x_2}\ket{1}\big) \xrightarrow{C-{R^*}_2}   
\\ 
\\
\qquad \qquad \qquad \qquad \frac{1}{\sqrt{2}}\big(\ket{0}+e^{2\pi i x_1*2^{-1} + x_2*2^{-2} -x_2*2^{-2} }\ket{1}\big) 
\\
\\
\qquad \qquad \qquad \qquad = \frac{1}{\sqrt{2}}\big(\ket{0}+e^{2\pi i x_1*2^{-1}}\ket{1}\big)
\\ 
\\ 
\qquad  \xrightarrow{H}  
\frac{1}{\sqrt{2}}\big(1+e^{2\pi i 0.x_1}\big)\ket{0} + \big(1-e^{2\pi i 0.x_1}\big)\ket{1} \\
\end{array}
\end{equation}

Calculating the probability from Eq. \ref{Cas2_twodigit} we have,

\begin{equation}\label{Probab_case2}
\begin{array}{l}
  P(\ket{0}) = \frac{1+cos(2\pi0.x_1)}{2}, \qquad
  P(\ket{1}) = \frac{1-cos(2\pi0.x_1)}{2}    
 \\   
\end{array} 
\end{equation}

The rotation gate $R_2$ is defined as Eq. \ref{Rotation_QFT} where $k=2$.  
\begin{equation}\label{Rotation_QFT}
   R_k = 
\begin{pmatrix}
    1       & 0 \\
    0       & e^{2\pi i/2^k} \\
\end{pmatrix}
\end{equation}


\subsection{Arbitrary constant precision phase estimation }
\label{constant-P}

This section follows the work describing an arbitrary precision QPE \cite{c25}. This approach reduces the number of shift operators for phase estimation and as a result decreases the depth of the quantum circuit. In this approach only the information from the two previous qubits are used to estimate the phase with constant precision. Controlled phase shift rotation $R_2$ and $R_3$ are applied to extract the information about the phase with arbitrary success probability. Fig.\ref{ACQPE} illustrates the circuit diagram for this arbitrary precision QPE approach.

\begin{figure}[t]
  \begin{center}
  \includegraphics[width=3.5in]{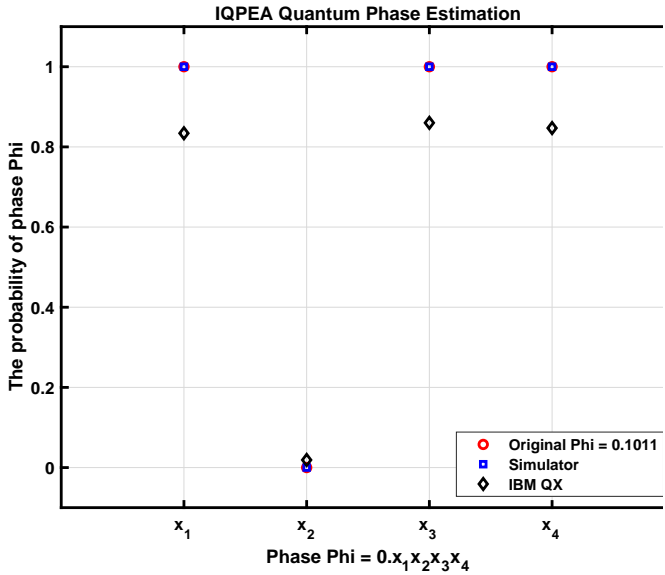}
  \caption{Iterative quantum phase estimation algorithm on Aer Simulator and IBMQX4}
  \label{iqpea}
  \end{center}
\vspace{-1.6em}
\end{figure}

The first stage of this approach is similar to QPE based on QFT. By applying the controlled gate $U^{2k}$ to the phase $\varphi=0.x_1x_2x_3...$, the state $\ket{\psi}$ will be given in Eq. \ref{ket_CP1}.

\begin{figure}[t]
  \begin{center}
  \includegraphics[width=3.5in]{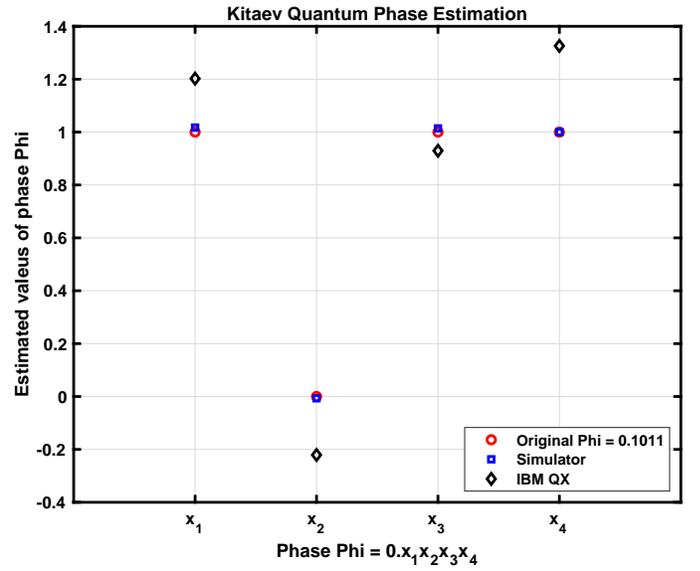}
  \caption{Kitaev quantum phase estimation algorithm on Aer simulator and IBMQX4}
  \label{kitaev}
  \end{center}
\vspace{-1.2em}
\end{figure}

\begin{equation}
\label{ket_CP1}
\ket{\psi_k} = \frac{1}{\sqrt{2}}\big(\ket{0}+e^{2\pi i *2^{k}\varphi}\ket{1}\big)
\\
\end{equation}

\begin{equation}
\label{ket_CP2}
\ket{y_i} = \frac{1}{\sqrt{2}}\big(\ket{0}+e^{2\pi i (0.x_i...x_n)}\ket{1}\big)
\\
\end{equation}

\noindent By applying controlled rotation $R_2^{-1}$ and $R_3^{-1}$ to the qubits and using the information from the two previous qubits we have,

\begin{equation}
\ket{\hat{\psi_k}} = \frac{1}{\sqrt{2}}\big(\ket{0}+e^{2\pi i *2^{k}\hat{\varphi}}\ket{1}\big)
\\
\end{equation}

\noindent where 

\begin{equation}
\hat{\varphi} = 0.x_{k+1} 0 0 x_{k+4}
\\
\end{equation}

\noindent Applying controlled rotation $R_2^{-1}$ and $R_3^{-1}$ will remove the effect of $x_{k+2}$ and $x_{k+3}$ so, the precision in this case will be,

\begin{equation}
|\varphi - 0.x_{k+1}| = \theta < \frac{1}{8}
\\
\end{equation}

\noindent Hence,

\begin{equation}
\ket{\hat{\psi_k}} = \frac{1}{\sqrt{2}}\big(\ket{0}+e^{2\pi i *2^{k} (0.x_{k+1}+\theta)} \ket{1}\big)
\\
\end{equation}

\noindent The post measurement probability based on the value of $\theta$ will be,

\begin{equation}
\begin{array}{l}
  P(0|k) = cos^2(\pi\theta) \geq cos^2(\frac{\pi}{8})\approx 0.85 ,\\
  \\
  P(1|k) = sin^2(\pi\theta) \leq sin^2(\frac{\pi}{8})\approx 0.15  
  \\
\end{array}
\end{equation}

As it can be seen only controlled rotation $R_2^{-1}$ and $R_3^{-1}$ are used in each stage to extract the estimated phase with $0.85$ success probability. Applying only two controlled rotation gates will reduce the number of operating gates and as a result will decrease the depth of the circuit. This improvement will help to implement the circuit using actual quantum computers and extract the phase with higher probability and less noise which reduces the estimation probability of the correct phase.

\section{Simulation and Experimental Results}
\label{sect-3}

In this section, we discuss the implementation of various quantum phase estimation (QPE) algorithms on both the Qiskit Aer simulator~\cite{c30} and the IBMQX4, IBM Q Experience 5-qubit quantum computing hardware platform~\cite{c22}. We obtained the theoretical results using the quantum simulator (Qiskit Aer) and then compared them with actual implementation on the QC hardware platform (IBMQX4).  It should be noted that the actual hardware measurements include all environmental errors within the system such as readout errors, gate errors and environmental noise. The inclusion of noise models in the simulators are beyond the specific work addressed here and will be investigated in future research and is address in the Section~\ref {discussion}.

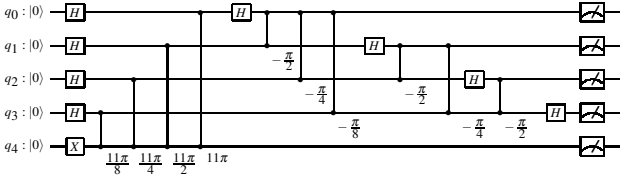
\begin{figure}[t]
 \begin{center}
 \tiny 
 \[\Qcircuit @C=.2cm @R=.2cm @!R {
         \push{\rule{4em}{0em}} &\lstick{q_0:\ket{0}}       & \gate{H} & \qw &\qw&\qw&\qw& \qw & \qw & \ctrl{4} &\qw & \gate{H}  & \ctrl{1} & \qw&\ctrl{2} & \qw & \ctrl{3} & \qw& \qw& \qw& \qw& \qw& \qw&\qw& \qw& \qw& \qw& \qw&\meter &\qw\\
         \push{\rule{4em}{0em}} &\lstick{q_1:\ket{0}}       & \gate{H} & \qw &\qw& \qw&\qw & \ctrl{3} &\qw &\qw & \qw &\qw& \ctrl{-1}&\dstick{-\frac{\pi}{2}}\qw &\qw&\qw &\qw&\qw &\gate{H} & \ctrl{1} \qw&\qw&\qw& \ctrl{2} &\qw& \qw& \qw& \qw& \qw& \meter &\qw \\
         \push{\rule{4em}{0em}} &\lstick{q_2:\ket{0}}       & \gate{H}  & \qw &\qw& \ctrl{2} &  \qw& \qw& \qw&  \qw& \qw& \qw& \qw&\qw& \ctrl{-2} &\dstick{-\frac{\pi}{4}}\qw&\qw& \qw& \qw& \ctrl{-1}&\dstick{-\frac{\pi}{2}}\qw& \qw&\qw& \gate{H} \qw& \ctrl{1}&\qw& \qw& \qw&  \meter &\qw \\
         \push{\rule{4em}{0em}} &\lstick{q_3:\ket{0}}       & \gate{H} & \ctrl{1}  & \qw& \qw& \qw& \qw& \qw& \qw& \qw&\qw& \qw&\qw&\qw&\qw& \ctrl{-3} &\dstick{-\frac{\pi}{8}}\qw& \qw& \qw& \qw& \qw & \ctrl{-2}& \dstick{-\frac{\pi}{4}}\qw& \ctrl{-1}&\dstick{-\frac{\pi}{2}}\qw& \qw&\gate{H} &\meter &\qw \\
        \push{\rule{4em}{0em}} &\lstick{q_4:\ket{0}}       &\gate{X}  & \ctrl{-1}  &\dstick{\frac{11\pi}{8}}\qw & \ctrl{-2} &\dstick{\frac{11\pi}{4}}\qw & \ctrl{-3} &\dstick{\frac{11\pi}{2}}\qw & \ctrl{-4} 
         &\dstick{11\pi}\qw &\qw& \qw& \qw& \qw& \qw& \qw& \qw&\qw& \qw& \qw& \qw& \qw& \qw&\qw& \qw& \qw& \qw&\meter &\qw\\ 
     }\]
\caption{Lloyd QPE algorithm gate with 1 ancillary qubit}
  \label{lloyd-orig-gate}
   \end{center}

\end{figure}


\begin{figure}[t]
  \begin{center}
  \includegraphics[width=3.5in,height=2.97in]{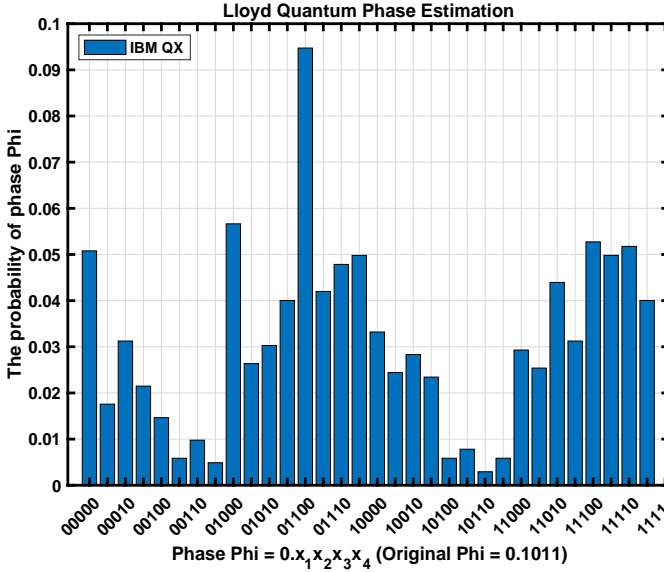}
  \caption{Lloyd QPE algorithm on IBMQX4}
  \label{lloyd-orig}
  \end{center}
\vspace{-1.67em}
\end{figure}

This work examined single qubit performance. The single qubit in IBM Q Experience has good fidelity on most quantum operations but the fidelity will be quickly degrade as the number of control qubits increases. Our results confirm that the accuracy of experimental results is significantly reduced as the number of qubits increases. To mitigate the problem, the modified solutions of these QPE algorithms were implemented in order to increase the accuracy of the phase that is being experimentally measured. The experimental procedures take advantages of the capability of classical computers to store intermediate results and then feed these values into the next quantum operation when appropriate. 

In our experiments, we set the phase $\varphi = 0.x_1x_2x_3x_4$, where the number of phase bit positions is 4 ($n=4$). We defined $\varphi=1/2 + 1/8 + 1/16$ which represents $\varphi=0.1011$ as a binary value. For each QPE algorithm, we ran the default $1,024$ shots for both simulator and the IBMQX4.

First, we implemented Kitaev's algorithm to find the phase $\varphi$ on both the Qiskit Aer simulator and IBMQX4 quantum computer. Fig.\ref{kitaev} shows that the estimated $\hat{\varphi}$ values of simulator results are almost the same as the original $\varphi$ values. The estimated $\varphi$ values from the IBM hardware platform are slightly different than the original $\varphi$ due to the lack of full quantum computing error correction capabilities today.


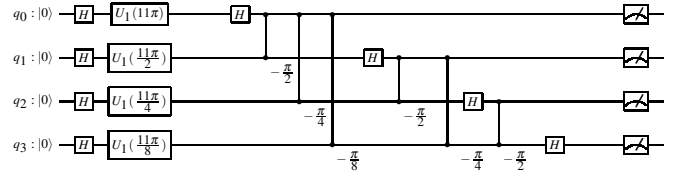
\begin{figure}[t]
 \begin{center}
 \tiny 
 \[\Qcircuit @C=.2cm @R=.2cm @!R {
         \push{\rule{4em}{0em}} &\lstick{q_0:\ket{0}}       & \gate{H} &\gate{U_1( 11\pi)}\qw &\qw &\qw &\qw & \gate{H}& \ctrl{1}& \qw&\ctrl{2} & \qw & \ctrl{3} & \qw& \qw& \qw& \qw& \qw& \qw&\qw& \qw& \qw& \qw& \qw& \qw&\qw&\qw&\meter &\qw\\
         \push{\rule{4em}{0em}} &\lstick{q_1:\ket{0}}       & \gate{H} &\gate{U_1( \frac{11\pi}{2})}&\qw &\qw & \qw &\qw& \ctrl{-1}&\dstick{-\frac{\pi}{2}}\qw &\qw&\qw &\qw&\qw &\gate{H} & \ctrl{1} \qw&\qw&\qw& \ctrl{2} &\qw& \qw& \qw& \qw& \qw& \qw& \qw&\qw&\meter &\qw \\
         \push{\rule{4em}{0em}} &\lstick{q_2:\ket{0}}       & \gate{H}  &\gate{U_1( \frac{11\pi}{4})}\qw &  \qw& \qw&\qw& \qw& \qw&\qw& \ctrl{-2} &\dstick{-\frac{\pi}{4}}\qw&\qw& \qw& \qw& \ctrl{-1}&\dstick{-\frac{\pi}{2}}\qw& \qw&\qw& \gate{H} \qw& \ctrl{1}&\qw& \qw& \qw& \qw& \qw&\qw& \meter &\qw \\
         \push{\rule{4em}{0em}} &\lstick{q_3:\ket{0}}       & \gate{H} & \gate{U_1( \frac{11\pi}{8})}\qw&\qw& \qw& \qw&\qw& \qw&\qw&\qw&\qw& \ctrl{-3} &\dstick{-\frac{\pi}{8}}\qw& \qw& \qw& \qw& \qw & \ctrl{-2}& \dstick{-\frac{\pi}{4}}\qw& \ctrl{-1}&\dstick{-\frac{\pi}{2}}\qw& \qw&\gate{H}\qw&\qw& \qw& \qw&\meter &\qw \\
     }\]
\caption{Modified Lloyd QPE algorithm gate without 1 ancillary qubit}
  \label{lloyd-gate}
   \end{center}
\end{figure}


\begin{figure}[t]
  \begin{center}
  \includegraphics[width=3.5in]{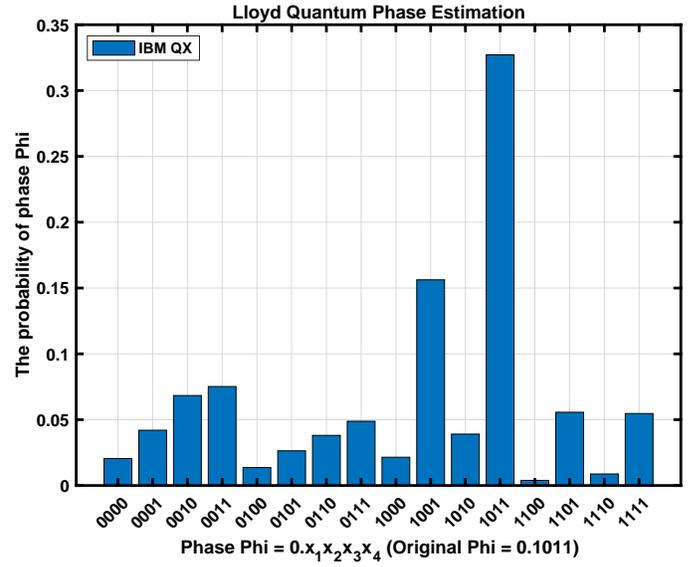}
  \caption{Modified Lloyd QPE algorithm on  IBMQX4}
  \label{lloyd}
  \end{center}
\vspace{-1.5em}
\end{figure}

Nevertheless, we can estimate the correct binary values of bit positions by converting the estimated $\varphi$ values. Because the noise can be attributed to various factors among different quantum computers, it is critical to find the hardware error rates in order to increase the accuracy of the $\varphi$ estimation in Kitaev's algorithm. The accuracy can be increased by adjusting proper error rates for each quantum computer during the computation process from the estimated $\varphi$ into the binary bit position. 

Second, we implemented iterative quantum phase estimation algorithm (IQPEA) to find the phase $\varphi$ on both the Qiskit Aer simulator and IBMQX4 quantum computer. Fig.\ref{iqpea} shows that the probability of finding $\varphi$ value from the simulation results are exactly the same as the original $\varphi$. The experiment results are slightly different than the original $\varphi$ but we can estimate the correct binary values the same way as Kitaev's algorithm.

Third, we implemented QPE algorithms using the inverse quantum Fourier transform technique to find the phase $\varphi$ on both the Qiskit Aer simulator and the IBMQX4 quantum computer. Fig.\ref{lloyd-orig} only shows the probability of finding $\varphi$ values from the IBMQX4 experiments because the probability of finding $\varphi$ value from the simulation results are exactly the same as the original $\varphi$. However, the highest probability of the phase $\varphi$ from the experimental results is when the $\varphi$ is $0.01100$ instead of the correct $\varphi=0.1011$.

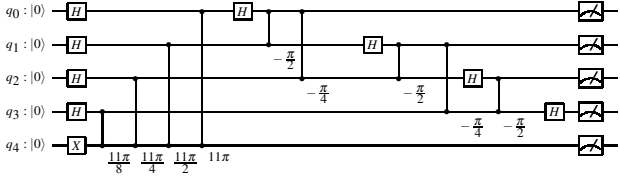
\begin{figure}[t]
 \begin{center}
 \tiny 
 \[\Qcircuit @C=.2cm @R=.2cm @!R {
         \push{\rule{4em}{0em}} &\lstick{q_0:\ket{0}}       & \gate{H} & \qw &\qw&\qw&\qw& \qw & \qw & \ctrl{4} &\qw & \gate{H}  & \ctrl{1} & \qw&\ctrl{2} & \qw & \qw & \qw& \qw& \qw& \qw& \qw& \qw&\qw& \qw& \qw& \qw& \qw&\meter &\qw\\
         \push{\rule{4em}{0em}} &\lstick{q_1:\ket{0}}       & \gate{H} & \qw &\qw& \qw&\qw & \ctrl{3} &\qw &\qw & \qw &\qw& \ctrl{-1}&\dstick{-\frac{\pi}{2}}\qw &\qw&\qw &\qw&\qw &\gate{H} & \ctrl{1} \qw&\qw&\qw& \ctrl{2} &\qw& \qw& \qw& \qw& \qw& \meter &\qw \\
         \push{\rule{4em}{0em}} &\lstick{q_2:\ket{0}}       & \gate{H}  & \qw &\qw& \ctrl{2} &  \qw& \qw& \qw&  \qw& \qw& \qw& \qw&\qw& \ctrl{-2} &\dstick{-\frac{\pi}{4}}\qw&\qw& \qw& \qw& \ctrl{-1}&\dstick{-\frac{\pi}{2}}\qw& \qw&\qw& \gate{H} \qw& \ctrl{1}&\qw& \qw& \qw&  \meter &\qw \\
         \push{\rule{4em}{0em}} &\lstick{q_3:\ket{0}}       & \gate{H} & \ctrl{1}  & \qw& \qw& \qw& \qw& \qw& \qw& \qw&\qw& \qw&\qw&\qw&\qw&\qw&\qw& \qw& \qw& \qw& \qw & \ctrl{-2}& \dstick{-\frac{\pi}{4}}\qw& \ctrl{-1}&\dstick{-\frac{\pi}{2}}\qw& \qw&\gate{H} &\meter &\qw \\
         \push{\rule{4em}{0em}} &\lstick{q_4:\ket{0}}       &\gate{X}  & \ctrl{-1}  &\dstick{\frac{11\pi}{8}}\qw & \ctrl{-2} &\dstick{\frac{11\pi}{4}}\qw & \ctrl{-3} &\dstick{\frac{11\pi}{2}}\qw & \ctrl{-4} 
         &\dstick{11\pi}\qw &\qw& \qw& \qw& \qw& \qw& \qw& \qw&\qw& \qw& \qw& \qw& \qw& \qw&\qw& \qw& \qw& \qw&\meter &\qw\\ 
     }\]
\caption{Arbitrary constant precision QPE gate with 1 ancillary qubit}
  \label{const-precision-orig}
   \end{center}
\end{figure}


\begin{figure}[t]
  \begin{center}
  \includegraphics[width=3.5in,height=2.97in]{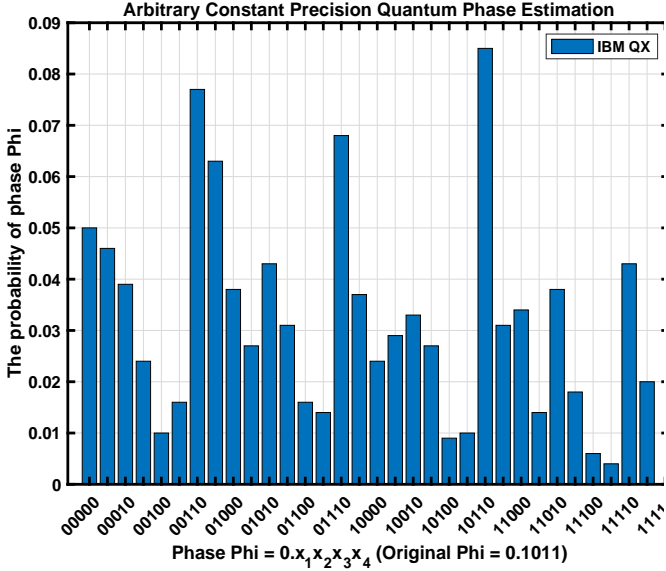}
  \caption{Arbitrary constant precision QPE on IBMQX4}
  \label{const-precision-orig}
  \end{center}
\vspace{-1.65em}
\end{figure}

The main reasons for these inaccurate results are caused by the lack of error correction capabilities, short longitudinal, and transverse coherence time for qubits and ancillary qubits respectively. Moreover, as described in Fig.\ref{lloyd-orig-gate}, the number of controlled phase rotation gates on qubits can increase the readout errors. 

To solve this problem and increase the accuracy of experimental results, we remove the ancillary control qubit and replace the unnecessary controlled-rotation gates with unitary rotation gates for each qubit as described in Fig.\ref{lloyd-gate}. Our experimental results Fig.\ref{lloyd} shows that our solution can find the correct phase $\varphi$ and even the probability (i.e., 0.335$\%$) is completely distinguished from other estimated $\varphi$ values.

Finally, we implemented the arbitrary constant precision (ACP) QPE algorithm based on the inverse quantum Fourier transform technique to find the phase $\varphi$ on both the Qiskit Aer simulator and the IBMQX4 quantum computer. Fig.\ref{const-precision-orig} only shows the probability of finding $\varphi$ values from the IBMQX4 experiments because the probability of finding $\varphi$ value from the simulation results are exactly the same as the original $\varphi$. However, the highest probability of the phase $\varphi$ from the experimental results is when the $\varphi$ is $0.10110$ instead of the correct $\varphi=0.1011$. 

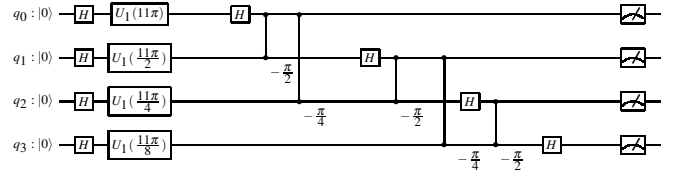
\begin{figure}[t]
 \begin{center}
 \tiny 
 \[\Qcircuit @C=.2cm @R=.2cm @!R {
         \push{\rule{4em}{0em}} &\lstick{q_0:\ket{0}}       & \gate{H} &\gate{U_1( 11\pi)}\qw &\qw &\qw &\qw & \gate{H}& \ctrl{1}& \qw&\ctrl{2} & \qw & \qw & \qw& \qw& \qw& \qw& \qw& \qw&\qw& \qw& \qw& \qw& \qw& \qw&\qw&\qw&\meter &\qw\\
         \push{\rule{4em}{0em}} &\lstick{q_1:\ket{0}}       & \gate{H} &\gate{U_1( \frac{11\pi}{2})}&\qw &\qw & \qw &\qw& \ctrl{-1}&\dstick{-\frac{\pi}{2}}\qw &\qw&\qw &\qw&\qw &\gate{H} & \ctrl{1} \qw&\qw&\qw& \ctrl{2} &\qw& \qw& \qw& \qw& \qw& \qw& \qw&\qw&\meter &\qw \\
         \push{\rule{4em}{0em}} &\lstick{q_2:\ket{0}}       & \gate{H}  &\gate{U_1( \frac{11\pi}{4})}\qw &  \qw& \qw&\qw& \qw& \qw&\qw& \ctrl{-2} &\dstick{-\frac{\pi}{4}}\qw&\qw& \qw& \qw& \ctrl{-1}&\dstick{-\frac{\pi}{2}}\qw& \qw&\qw& \gate{H} \qw& \ctrl{1}&\qw& \qw& \qw& \qw& \qw&\qw& \meter &\qw \\
         \push{\rule{4em}{0em}} &\lstick{q_3:\ket{0}}       & \gate{H} & \gate{U_1( \frac{11\pi}{8})}\qw&\qw& \qw& \qw&\qw& \qw&\qw&\qw&\qw& \qw& \qw\qw& \qw& \qw& \qw& \qw & \ctrl{-2}& \dstick{-\frac{\pi}{4}}\qw& \ctrl{-1}&\dstick{-\frac{\pi}{2}}\qw& \qw&\gate{H}\qw&\qw& \qw& \qw&\meter &\qw \\
     }\]
\caption{Arbitrary constant precision QPE gate without 1 ancillary qubit}
  \label{const-precision-orig}
   \end{center}
\end{figure}


\begin{figure}[t]
  \begin{center}
  \includegraphics[width=3.5in]{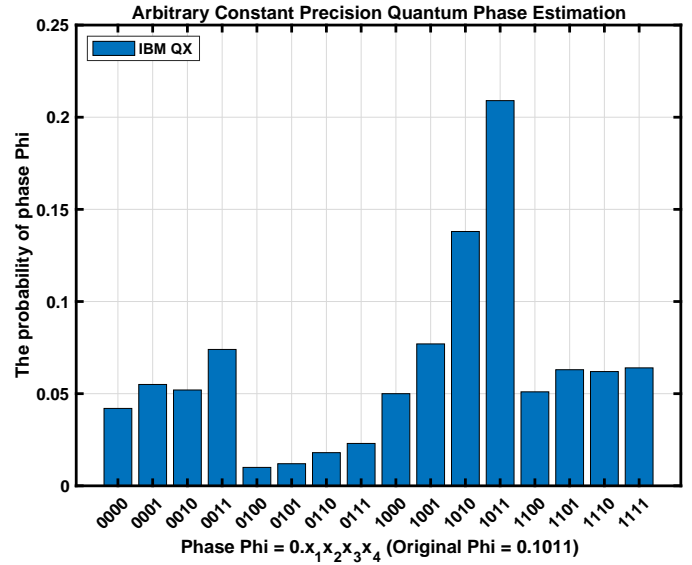}
  \caption{Modified arbitrary constant precision QPE on IBMQX4}
  \label{const-precision}
  \end{center}
\vspace{-1.5em}
\end{figure}

To increase the accuracy, the ancillary control qubit was removed and the unnecessary controlled-rotation gates were replaced with unitary rotation gates for each qubit as described in Fig.\ref{const-precision}. The experimental results shows that our solution can find the correct phase $\varphi$ and even the probability (i.e., 0.209$\%$) is completely distinguished from other estimated $\varphi$ values. However, the average accuracy of $i^{th}$ digit on ACP QPE algorithm is around $95\%$ so that the experimental results may vary with each experimental run.

\section{Discussion}
\label{discussion}
This paper investigates methods to increase the accuracy of implementing different QPE algorithms on actual quantum computers.  Although the paper presents several approaches to the QPE, this work mainly focuses on addressing practical challenges of implementing QPE based on the inverse QFT. The theoretical results of such QPE algorithms from the simulator estimate the phase with almost 100 $\%$ accuracy. However, it is not feasible to estimate the correct phase for QPE with inverse QFT from the NISQ quantum computers due to the noise of the system. 

The Kitaev approach can estimate the phase of an actual system using two qubits.  This approach provides a high fidelity and low error rate.  The disadvantage is that post-processing is required and there must be a relatively large number of measurements performed relative to the other methods investigated here for the determinations of the phase to be measured.

The inverse QFT method does not require post processing.  In addition, the binary digits of phase can be estimated separately.  However, the inverse QFT method requires a large number of rotation gates to achieve a precise solution.  The more gates in the system, the higher the level of noise.  Higher noise levels decrease the accuracy of finding the correct phase. 

The constant phase approach has the same set of advantages and disadvantages as the inverse QFT approach. However, one of the relative merits of this approach is that the number of required rotation gates of the original inverse QFT can be decreased by sacrificing some amount of the overall accuracy of the phase determination.

This paper showed that it was possible to remove the ancilla qubit from the  iterative QFT without the loss of its functionality, thereby removing the unnecessary controlled-rotation gates replacing them with unitary rotation gates for each qubit.  Making this change did reduce the number of controlled rotation gates required for a given level of accuracy.

All of these approaches were implemented on NISQ computers.  One of the properties of these machines is that there are multiple sources of measurement errors that do occur and can be attributed to the physical system  and the overall environment. For superconducting qubits coupled to readout cavities the state of the qubit is determined by measurement the response of a microwave tone incident on the readout cavity. Quantum computing hardware platforms do contain classical sources of noise that lead to readout errors of the qubit state. In addition it can also happen that not only the relaxation time $T_1$ decays the state of qubit during the measurement but also crosstalk between resonantors on chip and on the lines changes the probability distribution of the qubit states. These types of errors can be addressed by various techniques such as measurement calibration and error mitigation.

In this paper, the proposed approach was tested and analyzed using IBMQX4 which contains 5 qubits. The experimental results showed that using the proposed method the phase can be correctly estimated with reasonably distinct probability to other probabilities. The proposed approach can be easily applied to the large number of qubit systems to estimate the unknown phase of the complicated inputs. However, errors in the system described in this section can also increase as the number of qubits increases. Thus, it is critical that higher fidelity, longer coherence time, and lower readout errors should be followed by increasing the number of qubits to take full advantage of the presented technique.


\section{CONCLUSIONS}
\label{sect-4}
This paper demonstrates how to implement existing quantum phase estimation (QPE) algorithms on the state-of-the-art IBM quantum computers.  Our work also has documented the challenges of implementing QPE algorithms on real quantum processors.

We have proposed modified solutions of these algorithms by minimizing the number of controlled-rotation gates and by utilizing the digital computer's capabilities.  Our experimental results can guide researchers to consider these challenges when they implement their quantum algorithms on noisy intermediate-scale quantum (NISQ) computers.  Using these methodologies, substantial progress has been achieved applying QPE in various subject domains. 
 
The experimental results show that our solutions significantly increase the accuracy for finding correct phase.  Researchers can now implement these techniques using publicly available NISQ quantum computers such as the IBMQX4~\cite{c21} and Rigetti QPU~\cite{c22} in order to take better advantage of existing NISQ machines and advance toward the longer term goals of quantum advantage and ultimately quantum supremacy.




\end{document}